# Survey on Awareness of Privacy Issues in Ubiquitous Environment

Huma Tabassum, Sameena Javaid, Humera Farooq
Department of Computer Sciences
Bahria University, Karachi Campus
Karachi, Pakistan
humatabassum@rocketmail.com
sameenajaved@hotmail.com
humera.farooq@bimcs.edu.pk

*Abstract*—It is in human nature to keep certain details intimate. This confidential information, if made public, can result in annoyance, humiliation, and even devastating loss. The latter one holds true for individuals as well as organizations. In the case of ubiquitous environment, system adapts to the context of user's actions and intentions and the collection of data becomes pervasive. Thus, privacy turns out to be a vital aspect. This study aims to determine privacy awareness among people in ubiquitous environment. For this purpose, a survey was conducted. This survey was based on a questionnaire. The results show that people consider themselves quite aware, but, more concerned with privacy issues in ubiquitous environment. Also, a significant number of people admit to not taking privacy measures. The analysis was carried out on the basis of current status of participants in the university, namely, Undergraduate students, Graduate students, and Faculty members.

## I. INTRODUCTION

Ubiquitous Computing can be defined as "computing that is omnipresent and is, or appears to be, everywhere all the time; may involve many different computing devices that are embedded in various devices or appliances and operate in the background" [1]. The main focus of this paper is to determine privacy awareness in ubiquitous environment among people. For this purpose, a survey was conducted. The survey was conducted at Bahria University, Karachi campus during October-November, 2014. Participants were asked to fill out a questionnaire. Data samples were collected from students and faculty of various departments, which included Computer Sciences, Computer and Software Engineering, Electrical Engineering, Management Sciences, Humanities, Psychology, and Geophysics departments.

As ubiquitous environment masks the collection of information from the users, this study aims at analyzing awareness about privacy and its consequences. It also investigates the major privacy concerns of users as well as the measures they take. Legalization of privacy issues, improvement in privacy options, and any unpleasant experience of participants with respect to privacy issues, were also queried. In order to identify potential future dimensions in this area, level of comfort of participants along with the pros and cons of ubiquitous environment were also inquired.

The results indicate that majority of the participants consider themselves quite aware but more concerned with privacy issues; however, a significant number accepts that they do not take countermeasures for privacy issues.

The rest of this paper is organized as follows. The next section describes some of the work on privacy issues in existing research studies. Methodology for carrying out this study is discussed in the section after that. Then the results are discussed, followed by analysis. In the end, conclusion and future work is given which provides some directions to extend this study.

## II. RELATED WORK

Although ubiquitous environment provides ease and continuous access, it makes personal information widely available. Thus, privacy issues emerge as a major concern. Users are unknowingly providing their personal information to the environment which can be misused in case of any malicious attempt [2]. Every technology has both good and bad aspects; this implies that proper measures be taken to ensure privacy of users [15, 16]. It is, therefore, desired that users be aware of not only such systems and devices in their surroundings, but also of the issues these devices pose.

A recent study by Bonné, Quax, and Lamotte [3], shows that most Smartphone users are not truly aware of the privacy threats. Similarly, Renaud, Volkamer, and Renkema-Padmos [4] have also concluded that users are not aware of privacy protection. They tried to analyse mental model of users and made an effort to create understanding of the importance of addressing this issue. In their work, Acqusiti, John, and Loewenstein [5] tried to explore how important privacy is to the users. They argued on the rationale of the users and, on how their decisions affect their privacy by analysing the economic model and its effect on privacy issues.

In contrast, a survey based statistical analysis of Hoffman, Novak, and Peralta [6] revealed that approximate 69% of the web users refused to give their data or personal information online to any firm or individual, because they were not certain how their data can be used. Similarly, in another survey based study by Phelps, Novak, and Ferrell [7], 50% respondents claimed transparency about how organizations were using their personal information or individual-specific data. Several theoretical studies also defined that gender makes a great difference regarding privacy concerns while working online or on any social paltform [2].

As Angelelli [8] pointed out, major hurdles in providing a uniform method to ensure privacy include the constant advancements in technology, and that the privacy viewpoints differ among people. Thus, there cannot be a fixed number of ways to provide privacy. Usually, privacy issues arise due to improper configuration of a phone and also the network settings [3]. In order to facilitate the users, a recommenders system was proposed by Knijnenburg and Kobsa [9], to allow users to make decisions to disclose information in context.

For many researchers, revelation of the information or data in both aspects (context and use) is the primary fear. The major concerns related to privacy issues includestalking, unnecessary monitoring by employers or government agencies, and location, which can further lead to spam influx [10]. A study by Malhotra, Kim, and Agarwal [11] also presented some core privacy concerns. These were based on data collection and its control regarding privacy.

Privacy regarding personalization has been discussed in several studies [12]. In that study, personalization based on social connections and individual behaviors was analysed. It was shown to assist the programmers in resolving privacy issues. In another study carried out by Anton, Earp, and Young [13], it was revealed that privacy concerns of users have significantly been increasing since 2002.

In order to analyse the stated privacy concerns against the actual behaviour of people, Jiang, Heng, and Choi [14] have argued that people often do not act in accordance with their declared apprehensions. This is consistent with findings of this study that there are differences in what people say and what they actually do. Some new perspectives in online social communications and anonymity to ensure privacy were also discussed in their study[14].

The next section describes the methodology employed in carrying out this study.

### III. METHODOLOGY

In order to achieve the goals of this study, as mentioned in the beginning, a survey was conducted based on a questionnaire. The questions were designed to capture the essence of this study by exploring different approaches to gather the required information from participants. This study also aimed at identifying other potential dimensions related to privacy. These questions may be categorized in following ways.

#### A. General Context Questions

It captured the participants' awareness of ubiquitous environment by some 'feel-good' questions. These questions were helpful in establishing presence of the participants in ubiquitous environment, through careful inquiries of the activities they carry out using their smart device(s) and, their online habits. In addition, some other questions were also included in order to gather basic information, like gender, age group, and current status. These helped to categorize participants for performing analysis.

#### B. Key Privacy Questions

The survey then assessed level of both awareness and concern, of privacy issue in the participants' opinion through questions for each. Some other queries were intended to extract the major privacy concerns and issues of the participants, which came out to be personal information, financial information, and location. Participants were also inquired about the privacy measures they usually take (or not) along with the reasons for not taking any privacy measure. All these questions gave valuable insight to ascertain the behaviours and attitudes of participants with respect to privacy, which was the main objective of this study.

#### C. Trivial Privacy Questions

In order to understand if and why, privacy is a concerning issue, participants were asked about any unpleasant experience they might have encountered. Apart from this, opinion of participants on legalization of privacy issues was also queried. Participants were also inquired about how privacy options should be provided. This aided in establishing premise that such participants should be more conscious about privacy. However, as the results show, this was not always the case.

#### D. Other Dimensions

Finally, to generate further potential dimensions in future, a few questions about level of comfort of participants in ubiquitous environment, as well as the pros and cons of the ubiquitous environment, in general, were also included in the questionnaire. These dimensions can prove to be promising in future, if a thorough study is conducted to analyze the scope of the diverse aspects of privacy concerns and issues.

The next section discusses the results obtained from this study.

### IV. RESULTS

As described in the previous section, this study was based on survey in the form of a questionnaire. These questions were classified in four categories; namely general context questions, key and trivial privacy questions, and other dimensions. On the basis of this analysis criterion, the results obtained are discussed in the following sub-sections. It is important to note that these results reflect on participants' view and have been evaluated and compiled accordingly.

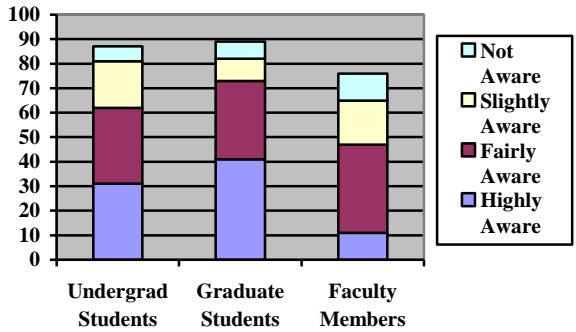

*Fig.1 Level of Privacy Awareness in each category of participants*

### A. Key Privacy Issues

*1) Privacy Awareness*: Among the participants who are part of ubiquitous environment, overall 71.4% considered themselves quite aware about privacy issues, and approximately 10%, not aware. The rest believed their selves somewhat aware of the privacy issues. Categorically, about 68% of the undergraduate students, 74% of graduate students, and 63% of the faculty members consider their selves to be quite aware.

Fig. 1 presents categorical distribution of participants in terms of their position in the university regarding privacy awareness.

*2) Privacy Concern:* After analysing awareness, this study revealed that participants are also concerned of the fact that their data can be used for malicious purpose provided it is easily accessible in ubiquitous environment. All the participants claimed to be concerned about privacy issues to varying extents. A relatively small group of participants (14%) stated to be slightly concerned, while others consider themselves quite concerned.

Fig.2 shows the concern of participants categorically. It should be noted that none of the participants declare that they are un-concerned about privacy.

Among those participants who claim to be concerned about privacy issues, around 80% were undergraduate students, 85% graduate students, and about 75% were faculty members.

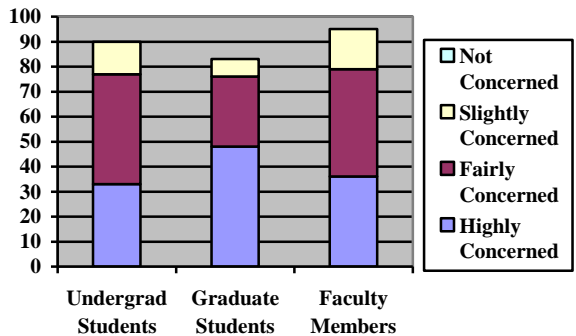

*Fig.2 Level of Privacy Concern in each category of participants*

TABLE I
COMPARISON OF OVERALL RESPONSES WITH EACH CATEGORY

|  | Overall Response | Categorized Response | | |
|---|---|---|---|---|
|  |  | *Undergraduate Students* | *Graduate Students* | *Faculty Member* |
| **Privacy Awareness** | 71.4% | 68% | 74% | 63% |
| **Privacy Concern** | 83.5% | 80% | 85% | 75% |

Table I encapsulates the responses of participants. Comparison of overall responses against each category is provided. It can be seen from the table that majority of the participants consider themselves quite aware as well as concerned about privacy issues. Another fact emphasized in the table is that the number of participants concerned with privacy issues, are more than those who are aware.

**Pressing Issues:** In this study, it was found that several participants are not only concerned about a single issue; they usually have apprehensions about more than one issue. The major concerns of participants were revealed to be their personal details including friends and family, location, and their financial information respectively.

A very high ratio of 94.5% participants is concerned about their personal details. Among these participants, about 47% were also concerned about their location, and for approximately 41%, their financial information was a concerning issue as well. For about 21% participants, all three of these issues were a privacy concern.

*3) Privacy Measures:* Dealing with privacy issues is another important aspect; however, in this survey, it was found that a considerable fraction of about 19% participants usually do not take any privacy measures. Major reasons for not taking measures are difficulties in configuration as per 65% of the participants. On the other hand, in some of the participants' view (about 4%); there is no need of taking any measures. Among the participants who take countermeasures, majority of about 64% admitted to disabling location with about 37% also modifying application settings with it. Other common privacy measure that approximately 29% participants took was giving false information.

Fig.3 depicts the categorical representation of participants on the basis of whether or not, they take privacy measures.

### B. Trivial Privacy Issues

Apart from the key issues analysed above, there were some trivial issues related to privacy that were also explored in this study. These queries provided valuable insight to understand the attitude of participants towards privacy issues.

One such aspect was related to legalization of privacy issues. Most of the participants (about 87%) believe that there should be laws to address privacy issues. However, only about 34% of the participants claimed to be aware of some existing laws.

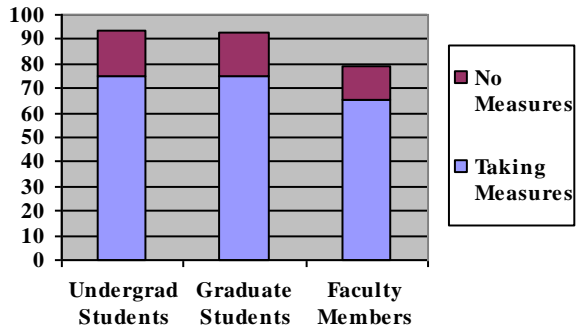

*Fig.3 Distribution among participants on the basis taking measures*

Among the 36% participants who admitted to have encountered some unpleasant experience due to their non-secure information, a significant 44% claimed to be a victim of fraud.

Another facet, in this study, analysed how the privacy settings can be improved. In the opinion of about 2% of the participants, there is no need for improvement. A very significant group of 57% demanded that more control be given to users over privacy settings.

*C. Other Dimensions*

Level of comfort of participants in ubiquitous environment is considered a future dimension in this study. It has been found that approximate 49% participants were quite comfortable with constant monitoring, while about 26% participants felt uncomfortable.

Another aspect, as a potential dimension for future was pros and cons of ubiquitous environment. While taking into account the merits of ubiquitous environment, the chief advantage was ease of life. Another advantage in the view of majority of the participants was fast access, whereas remote monitoring and socialization were also substantial benefits that the participants agreed upon in the survey.

After querying merits of ubiquitous environment, demerits were also questioned in the study. Theft of information and risk of exposure were the major apprehensions of participants, with 74.72% and 61.53% respectively. Other disadvantages in the view of participants were high energy consumption and increased machine reliance.

The next section describes the analysis strategy of this study. It discusses how the results were compiled and how the responses were evaluated.

V. ANALYSIS AND DISCUSSION

The main goals of this study were to evaluate privacy awareness among people in ubiquitous environment, their level of concern along with major issues, and what privacy measures they take. Apart from these major goals, this survey also investigated some other facets related to privacy as well as ubiquitous environment. For this purpose, data was collected at Bahria University, Karachi campus. The categorization of participants was done on the basis of their current status in university, that is, Undergraduate students, Graduate students, and Faculty members. The survey was conducted by taking samples of approximately one-fifth of each population in each category.

In order to achieve the above mentioned targets, first thing that needed to be established was whether the participants are part of ubiquitous environment or not. This was done on the basis of general context category of questions. These questions inquired the participants about the use of smart devices, smart appliances, and their usage habits. From the considerable data gathered, about 83% of the participants turned out to be part of ubiquitous environment. In terms of categorization, 86% were undergraduate students, 89% graduate students, and 68% belonged to faculty.

For every research objective, the responses of participants in each category, that is, undergraduate students, graduate students, and faculty, was evaluated separately. The overall percentage of responses has also been provided for comparison. The results were compiled in the light of these considerations.

The next section sums up the findings of this study in a conclusion.

VI. CONCLUSION

This study has revealed that a fair number of participants (71.4%) claim to be aware of the privacy issues in ubiquitous environment. However, the number of participants who declare their selves concerned about privacy issues, is more than those who claim to be aware (83.5%). Thus, it raises the question that how can people be concerned about something, which they are not aware of in the first place. In addition, despite claiming to be aware and concerned about privacy issues, a considerable number of participants (19%) admit to not taking any countermeasures. This leads to the conclusion that often, people act contrary to what they state.

The next section discusses future work which provides dimensions to extend the study.

VII. FUTURE WORK

For future work, the data may be analyzed on the basis of gender. This can generate an additional view in the privacy domain. Apart from this, other dimensions briefly mentioned in this study, can be explored further in future. These dimensions include comfort of participantsin ubiquitous environment and its pros and cons. These can be used to analyze the scope of the diverse aspects of privacy concerns and issues.

AUTHORS PROFILE

HUMA TABASSUM is currently pursuing her Masters degree in Computer Sciences from Bahria University, Karachi. She holds a Bachelors degree in Computer and Information Systems Engineering from N.E.D University of Engineering and Technology, Karachi. Her research interests include data engineering and ubiquitous computing environments.

SAMEENA JAVAID is currently doing her Masters in Computer Sciences from Bahria University, Karachi. She graduated in Computer Sciences from PIMSAT, Karachi. Her current research interests include theoretical foundations and e-learning with context-aware computing.

HUMERA FAROOQ is currently working as an Asst. Professor and Head of Computer Sciences department at Bahria University, Karachi. She did her PhD from Universiti Teknologi Petronas, Malaysia. Her research interests include machine learning, object tracking and detection, and smart environments.